\documentclass[12pt]{article}
\def\spireshome{http://www.slac.stanford.edu/cgi-bin/spiface/find/hep/www?FORMAT=WWW&}
{\catcode`\%=12
\xdef\spiresjournal#1#2#3{\noexpand\protect\noexpand\href{\spireshome
                          rawcmd=find+journal+#1%2C+#2%2C+#3}}
\xdef\spireseprint#1#2{\noexpand\protect\noexpand\href{\spireshome rawcmd=find+eprint+#1%2F#2}}
\xdef\spiresreport#1{\noexpand\protect\noexpand\href{\spireshome rawcmd=find+rept+#1}}
\xdef\spireskey#1{\noexpand\protect\noexpand\href{\spireshome key=#1}}
}
\def\eprint#1#2{\spireseprint{#1}{#2}{#1/#2}}

\def\nohref{}

\def\putpaper{\edef\refpage{\the\count0}%
              \def\nohref{}\def\onlyhref##1{}%
              {\def\ {+}\def\nohref##1{}\def\onlyhref{}%
               \edef\temp{\noexpand\spiresjournal
               {\journalname}{\volume}{\refpage}}\expandafter}\temp
               {\sfcode`\.=1000{\journalname} {\bf \volume}, 
                \refpage\ (\refyear)}\egroup}
\def\putpage{\edef\refpage{\the\count0}%
              \def\nohref{}\def\onlyhref##1{}%
              {\def\ {+}\def\nohref##1{}\def\onlyhref{}%
               \edef\temp{\noexpand\spiresjournal
               {\journalname}{\volume}{\refpage}}\expandafter}\temp
              {\refpage}\egroup}
\def\dojournal#1#2 (#3){\def\journalname{#1}\def\volume{#2}\def\refyear
                        {#3}\afterassignment\putpaper\bgroup\count0=}
\def\morepage{\afterassignment\putpage\bgroup\count0=}

\def\NPB#1{\dojournal{Nucl.\ Phys.}{B#1}}
\def\NPBPS#1{\dojournal{Nucl.\ Phys.\ \nohref{{\bf B }}\nohref(Proc.\ Suppl.\nohref)}{#1}}

\def\PRD#1{\dojournal{Phys.\ Rev.\nohref{ D }}{\onlyhref D#1}}

\def\PLB#1{\dojournal{Phys.\ Lett.}{B#1}}

\usepackage[hyperindex]{hyperref}
\pagestyle{plain}
\input{epsf}

\setlength{\topmargin}{-2em}
\setlength{\oddsidemargin}{-1ex}
\addtolength{\textwidth}{17ex}
\addtolength{\textheight}{8em}

\newcommand{\delv}{{\bf \nabla}}
\newcommand{\delvc}{{\bf D}}

\newcommand{\delfour}{{\Delta^{(4)}}}
\newcommand{\delsq}{\Delta^{(2)}}

\newcommand{\Sigmav}{\mbox{\boldmath$\Sigma$}}
\newcommand{\gammav}{\mbox{\boldmath$\gamma$}}
\newcommand{\alphav}{\mbox{\boldmath$\alpha$}}

\newcommand{\Mbz}{{M_0}}

\newcommand{\be}{\begin{equation}}
\newcommand{\ee}{\end{equation}}
\newcommand{\order}{{\cal O}}
\newcommand{\lag}{{\cal L}}

\newcommand{\nl}{\nonumber \\}

\newcommand{\Ev}{{\bf E}}
\newcommand{\Bv}{{\bf B}}

\newcommand{\sigmav}{\mbox{\boldmath$\sigma$}}
\newcommand{\ainv}{$a^{-1}$}
\newcommand{\ainvsp}{$a^{-1}\;$}

\begin{document}

\ifx\href\undefined\global\def\href#1#2{{#2}}\fi

\title{ \vspace{-1cm}\begin{flushright}
{\small OHSTPY-HEP-T-97-017 \\
 LAUR-97-4642 \\
UCSD/PTH 98-12 \\}
\end{flushright}
{
\vspace{1cm}
\bf B Meson Decay Constants From NRQCD 
}}

\author{
{\bf A.Ali Khan$^a$, T.Bhattacharya$^b$, S.Collins$^c$,
C.T.H.Davies$^c$, R.Gupta$^b$,} \\  {\bf C.Morningstar$^d$, 
J.Shigemitsu$^a$, J.Sloan$^e$.}\\[.6cm]
\small $^a$Physics Department, The Ohio State University,\\ 
\small Columbus, OH 43210, USA.\\[.2cm]
\small $^b$T-8, Los Alamos National Laboratory,\\
\small  Los Alamos, NM 87545, USA.\\[.2cm]
\small $^c$Department of Physics \& Astronomy, University of Glasgow, \\
\small Glasgow, UK G12 8QQ. \\[.2cm]
\small $^d$Physics Department, University of California at San Diego, \\
\small La Jolla, CA 92093, USA. \\[.2cm]
\small $^e$Physics Department, University of Kentucky,\\
\small Lexington, KY 40506, USA.
\\ }

\date{May 1998}

\maketitle

\begin{abstract}
\noindent
We present quenched results for $B$ meson decay constants using NRQCD 
$b$ quarks and \order($a$) tadpole improved clover light quarks. 
 For the first time, one-loop matching factors between lattice and 
continuum currents are 
incorporated through \order($\alpha/M$) taking 
 operator mixing fully into account. 
This includes an important \order($\alpha \, a$) discretization correction 
 to the heavy-light axial vector current.
We find $f_B = 147(11)(^{+8}_{-12})(9)(6)$MeV and 
$f_{B_s}/f_B = 1.20(4)(^{+4}_{-0})$.

 \vspace{.1in}

\end{abstract}

\section{Introduction}

The $B$ meson decay constant $f_B$ and the bag parameter $B_B$ play
crucial roles in analyses of $B_0 - \overline{B_0}$ mixing phenomena
and the extraction of CKM matrix elements.  Experimental determination
of $f_B$ is severely constrained by the very small leptonic branching
fraction, while $B_B$ is not a direct outcome of experiments. Their
calculation has proven difficult due to the possibility of large QCD
corrections to the basic weak process.  Lattice QCD provides a
non-perturbative method for calculating the relevant hadronic matrix
elements from first principles \cite{review}.  This paper presents a
quenched lattice calculation of $f_B$ and $f_{B_s}$ using the
non-relativistic QCD (NRQCD) approach to $b$-quarks~\cite{cornell}.
Much of our discussion of the systematic improvement of 
$f_B$, however, applies to other lattice calculations of this quantity.  

In the rest frame of the $B$ meson, the momentum of the $b$-quark 
 is of \order($\Lambda_
{QCD}$).  The $b$-quark velocity is hence  $v \approx \Lambda_{QCD}/M 
\approx 0.1$, where $M$ is the heavy quark mass, 
and the nonrelativistic approach  is sensible. 
  The heavy quark action that we employ includes relativistic corrections
 through
\order($(\Lambda_{QCD}/M)^2$). 
 For the light quarks we use the tree-level tadpole
improved clover action which reduces \order$(a\Lambda_{QCD})$ errors to 
\order$(\alpha \,(a \Lambda_{QCD}))$ \cite{clover}.
In the heavy-light axial current we investigate both $1/M$ and $1/M^2$
corrections.  The one-loop matching factors relating the lattice and
full (relativistic) continuum QCD currents have been calculated to
\order$(\alpha/M)$~\cite{pert1} and incorporated in this analysis. 
Mixing among heavy-light current operators and contributions from 
 an \order($ \alpha (a \Lambda_{QCD})$) 
discretization  correction to the current are taken into account.  
In the nonrelativistic approach the $1/M$ expansion and \order$(a)$ 
improvements go hand in hand.
Previous calculations of $f_B$ using NRQCD $b$-quarks have included 
only subsets of the above ingredients \cite{othernrqcd,stlouis,hiroshima}, 
as have calculations using the relativistic formalism for the $b$-quark 
\cite{APE,fermilab,JLQCD,MILC,UKQCD}. 
 Here we include 
 all of them and carry out a comprehensive study with improvements
 through \order($(\Lambda_{QCD}/M)^2$) 
and \order($a \Lambda_{QCD}$) in the action and through 
\order($(\Lambda_{QCD}/M)^2$), \order($\alpha/(aM)$), \order($\alpha 
(\Lambda_{QCD}/M)$) 
and \order($\alpha (a \Lambda_{QCD})$) in the currents.

\section{Details of the Simulation }

\vspace{.2in}
\noindent
The NRQCD heavy quark action density used in our lattice simulations is
a variant of the one proposed in \cite{cornell},
 \be \label{nrqcdact}
 \lag =  \overline{\psi}_t \psi_t - 
 \overline{\psi}_t
\left(1 \!-\!\frac{a \delta H}{2}\right)_t
 \left(1\!-\!\frac{aH_0}{2n}\right)^{n}_t
 U^\dagger_4
 \left(1\!-\!\frac{aH_0}{2n}\right)^{n}_{t-1}
\left(1\!-\!\frac{a\delta H}{2}\right)_{t-1} \psi_{t-1},
 \ee
where $\psi$ denotes a two-component Pauli spinor, 
 $H_0$ is the nonrelativistic kinetic energy operator,
 \be
 H_0 = - {\delsq\over2\Mbz},
 \ee
and $\delta H$ includes relativistic and finite-lattice-spacing
corrections,
 \begin{eqnarray}
\delta H 
&=& - \frac{g}{2\Mbz}\,\sigmav\cdot\Bv \nl
& & + \frac{ig}{8(\Mbz)^2}\left(\delv\cdot\Ev - \Ev\cdot\delv\right)
 - \frac{g}{8(\Mbz)^2} \sigmav\cdot(\delv\times\Ev - \Ev\times\delv)\nl
& & - \frac{(\delsq)^2}{8(\Mbz)^3} 
  + \frac{a^2\delfour}{24\Mbz}  - \frac{a(\delsq)^2}{16n(\Mbz)^2} \;.
\label{deltaH}
\end{eqnarray}
 In addition to all $1/M^2$ terms we have also included the leading 
$1/M^3$ relativistic correction 
as well as the discretization corrections appearing at the same 
order in the momentum expansion. 
 $\delv$ and $\delsq$ are the symmetric gauge-covariant lattice derivative
and laplacian, while $\delfour$ is a discretized version of the continuum
operator $\sum D_i^4$. The parameter $n$ is introduced to remove 
instabilities in the heavy quark propagator due to the highest momentum 
modes~\cite{cornell}. For the heavy quark action,  we use two-leaf $\Ev$ 
fields~\cite{fermilabaction} rather than the four-leaf 
fields of previous NRQCD simulations.
The tree-level coefficients are tadpole-improved by 
rescaling the gauge fields by the fourth root of the 
plaquette,  $u_0 = 0.87779$ at $\beta = 6.0$
 \cite{lepmac}. 
For the light quarks we use the tree-level tadpole improved clover action
 ($c_{SW} = u_0^{-3}$).

Simulations were carried out on 102 quenched configurations on $16^3
\times 48$ lattices at $\beta = 6.0$. Both forward and time reversed
quark propagators were calculated on each configuration in order to
improve statistics.  Error estimates are obtained using a bootstrap
procedure. Meson
correlators are calculated for 30 combinations of quark masses; five
values for $\kappa$ and six values for the bare heavy quark mass.  Two
of the $\kappa$ values are selected to simulate the charm quark, and
the remaining three are around 
the strange quark mass.  Heavy quark masses range from below the bottom 
quark mass to about five times the bottom quark mass. 
 Both the heavy and the light quark propagators 
were smeared at the sink and/or the source. 
 Further details are given in Ref.~\cite{stlouis}. 

There are four input parameters in the simulations, the light quark
mass ($\kappa_l \equiv \kappa_{u,d}$), the strange quark mass ($\kappa_s$), 
the scale \ainvsp, and the heavy quark mass ($M_0$).  These need to be
fixed via experimental input. 

To set the scale one can use a variety of quantities like $M_\rho$,
$f_\pi$, string tension ($\sigma$), or quarkonium S--P splittings. In 
quenched calculations, however, 
different physical quantities 
 can lead to different \ainvsp's   and one must make a choice. 
 In principle one would like to
choose an observable whose 
 characteristic momenta are similar to those of the quantities being 
calculated 
and which has comparable quenching  and discretization errors.
The first choice would be to  fix $a^{-1}$ from the 
heavy-light spectrum 
itself but experimental and numerical errors will have to be reduced before
 this becomes practical.  
 Based on the ``brown muck'' picture of HQET for heavy-light 
systems~\cite{isgur},  however, we believe 
 that the scales relevant for energy splittings in heavy-light spectroscopy 
and for the decay constant are determined by the light quark dynamics.
For $f_B$ and spectroscopy 
 we shall therefore consider observables in the light quark and gluonic
sector to set the scale.
 For clover fermions at $\beta=6.0$, the values for $a^{-1}$
from $M_\rho$, $M_K$, $M_N$, $M_\Delta$, $\sigma$, and $f_\pi$ range
between $1.8$ GeV and $2.0$ GeV. We use $a^{-1}_{m_\rho} = 1.92(7)$
GeV as it is close to the mean, and can be extracted most reliably
from our data. We also calculate the final values using $a^{-1} = 1.8$ and 
 $2.0$ GeV and use the spread as the uncertainty due to the 
determination of the scale.

 We determine $\kappa_l = 0.13917(9)$ by linearly
extrapolating the pion mass, calculated on the same set of
configurations, to the physical value. To set $\kappa_s$ we use  
the $K$, the $K^*$ and the $\phi$, which yields the values 0.13755(13),
0.13719(25), and 0.13717(25), respectively.  We quote central values for our 
$B_s$ meson 
decay constant using the $K$, and the difference between the $K$
and $\phi$ as a systematic error.

\vspace{.1in}
\noindent
The bare heavy quark mass is fixed by linearly interpolating between the meson
masses with the three lightest $M_0$ values to the experimental
 $B$ meson mass.
Since the NRQCD action omits the 
heavy quark rest mass,  heavy-light pseudoscalar meson masses are 
 not direct outputs of the simulations.  Instead,  correlation 
functions fall off with an energy $ E_{sim}$ that is related to 
the full meson mass $M_{PS}$ and the bare $b$-quark mass $M_0$ as,
\be  \label{mps}
M_{PS} = \Delta + E_{sim} \equiv 
\left[ Z_m\;M_0 - E_0 \right] + E_{sim},
\ee
where $Z_m$ is the mass renormalization constant and $E_0$ an 
energy shift. 
  $Z_m$ and $E_0$ have been calculated perturbatively and 
$E_{sim}$ is obtained nonperturbatively from the simulations.
An alternate way to obtain the meson mass is to calculate energy splittings 
between correlators with and without spatial momentum,
\be  \label{mkin}
E_{sim}(\vec{p}) - E_{sim}(0) = \sqrt{\vec{p}^2 + M_{kin}^2} - M_{kin} .
\ee
In Table~\ref{tab:masses} we list results for $M_{PS}$ of eq.(\ref{mps}) 
using perturbation 
theory to calculate $\Delta$ (denoted ``$M_{pert}$'' in the Table) 
and for $M_{kin}$ 
of eq.(\ref{mkin}), both  measured in units of 
\ainvsp. 
 $\Delta$ can also be obtained nonperturbatively from 
simulations of quarkonium $\overline{Q}Q$ systems, whose correlations fall 
off with an energy $E_{sim}^{\overline{Q}Q}$ related to the full 
$\overline{Q}Q$ meson mass as,
\be  \label{mqbarq}
M_{\overline{Q}Q} =
M^{\overline{Q}Q}_{kin} 
= 2\,\Delta + E_{sim}^{\overline{Q}Q}.
\ee
Both  $M^{\overline{Q}Q}_{kin} $ and  $E_{sim}^{\overline{Q}Q}$ can be 
extracted from the simulations, leading to a nonperturbative determination 
of $\Delta$.  The resulting masses for the heavy-light mesons are also listed 
in Table~\ref{tab:masses} under $ aM'$. 
 One finds consistency among the three determinations of
 the meson mass,
$aM_{kin}$, $aM_{pert}$ and $aM'$, in most cases, to within $1 \sigma$. 
 The errors for $aM_{kin}$ are much larger than for 
$aM'$ since finite momentum correlators have much larger fluctuations 
for heavy-light mesons than for heavy-heavy mesons.  
The errors on $aM_{pert}$ 
 were estimated by squaring the one-loop 
corrections  and by varying $q^*$, 
the scale in the coupling $\alpha$. 
 We will use the most accurately determined 
$aM'$ to fix $aM_0$. 
With \ainv = 1.92(7) GeV and using linear interpolation between the three 
lightest heavy quark masses,  one finds
 that $aM_0 = 2.22(11)$ gives the 
correct $B$ meson mass.  Had one used $aM_{kin}$ 
rather than $aM'$, the number changes to 
$aM_0 = 2.28(42)$.  The effect of this difference 
 on $f_B$ is invisible.

\begin{table}
\begin{center}
\begin{tabular}{|l|l|l|l|l|l|l|l|l|}
\hline
\multicolumn{1}{|c}{} &
\multicolumn{4}{|c|}{$\kappa_{l}$} &
\multicolumn{4}{|c|}{$\kappa_{s}$} \\
\hline
\multicolumn{1}{|c}{$aM_0$} &
\multicolumn{1}{|c}{$aE_{sim}$} &
\multicolumn{1}{|c}{$aM_{kin}$} &
\multicolumn{1}{|c}{$aM_{pert}$} &
\multicolumn{1}{|c|}{$aM'$} &
\multicolumn{1}{|c}{$aE_{sim}$} &
\multicolumn{1}{|c}{$aM_{kin}$} &
\multicolumn{1}{|c}{$aM_{pert}$} &
\multicolumn{1}{|c|}{$aM'$} \\
\hline
1.6  & 0.422(6)& 2.17(24)   &2.08(09) & 2.11(5) 
& 0.483(8) & 2.21(13)   &2.14(09) & 2.17(5)\\  
2.0  & 0.438(7)& 2.57(33)   &2.46(10) & 2.52(6)
& 0.499(8)& 2.63(18)   &2.52(10) & 2.58(6)\\  
2.7  & 0.452(8)& 3.34(58)   &3.12(11) & 3.26(8)
& 0.513(8)& 3.37(29)   &3.18(11) & 3.32(8)\\  
4.0  & 0.461(9)& 5.2(1.5)  &4.32(10)  & 4.59(9)
 & 0.522(8)   & 4.86(67)  &4.38(10) & 4.65(9) \\         
\hline
\end{tabular}
\end{center}
\caption{Simulation energies and pseudoscalar meson masses 
in lattice units for light and 
strange clover quark masses. Meson masses are determined from 
 eq.\protect(\ref{mps}) with $\Delta$ coming 
from perturbation theory ($M_{pert}$) or  
heavy-heavy spectroscopy ($M'$), and from the dispersion relation 
eq.\protect(\ref{mkin}) ($M_{kin}$).
}
\label{tab:masses}
\end{table}

The results for the heavy-light spectrum,
including radially excited states, P-states and heavy baryons, based
on the same data set, are summarized in Figure 1. 
  We find that the combination, NRQCD heavy and clover light
quarks, reproduces the gross features of the $B$ mesons and the heavy
baryons. This is a prerequisite for a reliable calculation of $f_B$.
Details of the spectrum calculations will be given in
a separate publication \cite{hlspec}.

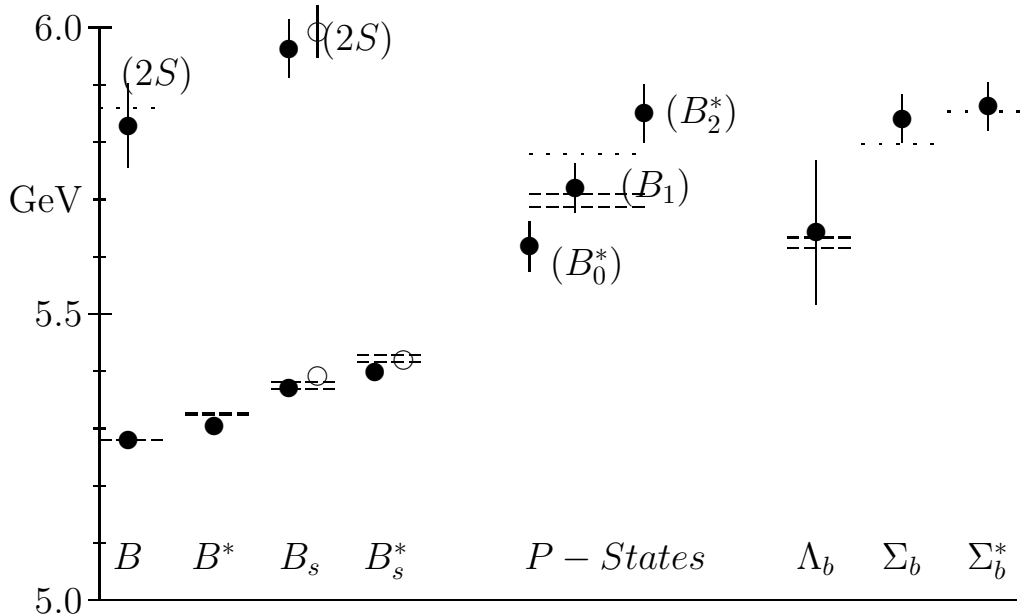
\begin{figure}
\begin{center}
\setlength{\unitlength}{.030in}
\begin{picture}(130,100)(30,500)
\put(15,500){\line(0,1){100}}
\multiput(13,500)(0,50){3}{\line(1,0){4}}
\multiput(14,500)(0,10){11}{\line(1,0){2}}
\put(12,500){\makebox(0,0)[r]{{\large5.0}}}
\put(12,550){\makebox(0,0)[r]{{\large5.5}}}
\put(12,600){\makebox(0,0)[r]{{\large 6.0}}}
\put(12,570){\makebox(0,0)[r]{{\large GeV}}}
\put(15,500){\line(1,0){160}}


     \put(20,510){\makebox(0,0)[t]{{\large $B$}}}
     \put(20,528){\circle*{3}}
     \multiput(15,527.9)(3,0){4}{\line(1,0){2}}
     \put(20,582.9){\circle*{3}}
     \put(20,582.9){\line(0,1){7.3}}
     \put(20,582.9){\line(0,-1){7.3}}
     \put(25,595){\makebox(0,0)[t]{{\large $(2S)$}}}
     \multiput(15,586)(3,0){4}{\line(1,0){0.5}}

     \put(35,510){\makebox(0,0)[t]{{\large $B^{*}$}}}
     \put(35,530.5){\circle*{3}}
     \multiput(30,532.6)(3,0){4}{\line(1,0){2}}
     \multiput(30,532.4)(3,0){4}{\line(1,0){2}}
     \put(50,510){\makebox(0,0)[t]{{\large $B_s$}}}
     \put(48,537.1){\circle*{3}}
     \put(48,537.1){\line(0,1){1}}
     \put(48,537.1){\line(0,-1){1}}
     \put(53,539.2){\circle{3}}
     \multiput(45,538.1)(3,0){4}{\line(1,0){2}}
     \multiput(45,536.9)(3,0){4}{\line(1,0){2}}
     \put(48,596.3){\circle*{3}}
     \put(48,596.3){\line(0,1){5.1}}
     \put(48,596.3){\line(0,-1){5.1}}
     \put(53,599.3){\circle{3}}
     \put(53,599.3){\line(0,1){4.6}}
     \put(53,599.3){\line(0,-1){4.6}}
     \put(60,601){\makebox(0,0)[t]{{\large $(2S)$}}}

     \put(65,510){\makebox(0,0)[t]{{\large $B^{*}_s$}}}
     \put(63,539.8){\circle*{3}}
     \put(63,539.8){\line(0,1){1.1}}
     \put(63,539.8){\line(0,-1){1.1}}
     \put(68,541.9){\circle{3}}
     \multiput(60,542.8)(3,0){4}{\line(1,0){2}}
     \multiput(60,541.6)(3,0){4}{\line(1,0){2}}
     \put(105,510){\makebox(0,0)[t]{{\large $P-States$}}}
     \put(110,585){\circle*{3}}
     \put(110,585){\line(0,1){5}}
     \put(110,585){\line(0,-1){5}}
     \put(120,585){\makebox(0,0){{\large $(B^*_2)$}}}
     \put(90,561.8){\circle*{3}}
     \put(90,561.8){\line(0,1){4.4}}
     \put(90,561.8){\line(0,-1){4.4}}
     \put(100,561.8){\makebox(0,0)[t]{{\large $(B^*_0)$}}}
     \put(98,572){\circle*{3}}
     \put(98,572){\line(0,1){4.2}}
     \put(98,572){\line(0,-1){4.2}}
     \put(112,572){\makebox(0,0){{\large $(B_1)$}}}
     \multiput(90,568.6)(3,0){7}{\line(1,0){2}}
     \multiput(90,571.0)(3,0){7}{\line(1,0){2}}
     \multiput(90,577.9)(3,0){7}{\line(1,0){0.5}}

     \put(140,510){\makebox(0,0)[t]{{\large $\Lambda_b$}}}
     \put(140,564.2){\circle*{3}}
     \put(140,564.2){\line(0,1){12.5}}
     \put(140,564.2){\line(0,-1){12.5}}
     \multiput(135,561.5)(3,0){4}{\line(1,0){2}}
     \multiput(135,563.3)(3,0){4}{\line(1,0){2}}

     \put(155,510){\makebox(0,0)[t]{{\large $\Sigma_b$}}}
     \put(155,584.1){\circle*{3}}
     \put(155,584.1){\line(0,1){4.2}}
     \put(155,584.1){\line(0,-1){4.2}}
     \multiput(148,579.7)(3,0){5}{\line(1,0){0.5}}

     \put(170,510){\makebox(0,0)[t]{{\large $\Sigma_b^*$}}}
     \put(170,586.2){\circle*{3}}
     \put(170,586.2){\line(0,1){4.2}}
     \put(170,586.2){\line(0,-1){4.2}}
     \multiput(163,585.3)(3,0){5}{\line(1,0){0.5}}

\end{picture}
\end{center}
\caption{Heavy-light spectrum with $b$-quarks.  Horizontal dashed lines 
bracket experimental values taken from the Particle Data Book. 
Horizontal dotted lines indicate first experimental observations of 
these states (\protect\cite{lep1} (2S \& $1^{+ \prime}/B^*_2$) and 
\protect\cite{lep2} ($\Sigma_b$)). 
 Errors on simulation results include statistical plus, where appropriate,
errors from extrapolation 
 in the light quark mass to the physical pion. The two results 
for the strange $B$ mesons correspond to fixing the strange quark mass 
from either the $K$ (filled circles) or the $\phi$ (open circles). 
For the $J=1$  P states only one level is shown, since we are currently 
unable to reliably separate 
 the physical $1^+$ and $1^{+\prime}$ states.
}
\end{figure}

\nopagebreak
\section{Heavy-Light Axial Currents in  NRQCD and Matching to 
Continuum Full QCD}

\vspace{.1in}
\noindent
The goal is to calculate the pseudoscalar decay constant $f_{PS}$,
defined in Euclidean space through the relation,

\be \label{deffb}
   \langle \, 0 \,| \, A_{\mu} \,|\,PS\, \rangle = p_{\mu} f_{PS} . 
\ee
Here $A_{\mu}$ is the heavy-light axial vector current in full continuum 
QCD.  One needs to relate this current to the current operators $J^{(i)}_L$ 
in lattice NRQCD.

\vspace{.1in}
\noindent
Heavy-light currents in full continuum QCD have the form 
 $\bar{q} \Gamma h$. For $A_0$, we use
$\Gamma = \gamma_5 \gamma_0$.   The 
four-component Dirac spinor for the heavy quark, $h$, is related to the 
two-component NRQCD heavy quark (heavy anti-quark)
fields, $\psi$ ($\tilde{\psi}$), via an inverse Foldy-Wouthuysen 
transformation:

\be \label{fw}
h = U^{-1}_{FW} \Psi_{FW} = U^{-1}_{FW} \left(\begin{array}{c} 
                                        \psi  \\
                                      \tilde{\psi}
                       \end{array}  \right).
\ee

\noindent
Through \order($ 1/M^2$) this gives, 
\begin{eqnarray} \label{jtree}
\bar{q} \Gamma h  &=& \bar{q} \Gamma Q - \frac{1}{2 M}\, \bar{q}\,
 \Gamma (\gammav \cdot \delvc) Q  \nl
 & & + \frac{1}{8 M^2}\,\bar{q}\,\Gamma\left(\delvc^2 + g \Sigmav
 \cdot \Bv -2ig \alphav \cdot \Ev \right) \,Q,
\end{eqnarray}
where $Q = \frac{1}{2}(1 + \gamma_0) \Psi_{FW} $ and 
$\alphav \equiv \gamma_0 \gammav$,  $\Sigmav = 
diag(\sigmav, \sigmav)$. 
In our simulations we have included all tree-level contributions to the 
 current through \order($1/M^2$) 
using the terms in eq.(\ref{jtree}).

\vspace{.1in}
\noindent
 We use perturbation theory to match between the lattice currents  in
our simulations and those of continuum QCD and employ
 on-shell matching conditions.  
 One starts by considering the process in which a
 heavy quark of momentum $p$ is scattered by the heavy-light current into  
a light quark of momentum $ p^{\prime}$.  Matching takes place by 
requiring that  
the amplitude for this process calculated in full QCD agrees with that in 
the effective theory through the order in $\alpha$ and $1/M$ that one 
is working in. Here we  work through \order($\alpha/M$). 
The amplitude in full QCD 
 can be expanded in terms of $p/M$, $p^{\prime}/M$ etc.  
The next step is to identify current operators in the effective theory 
that would reproduce the same terms in the amplitude.  Finally, a 
one-loop mixing matrix calculation must be done within the effective 
theory.  Details of such a matching calculation are given in \cite{pert1}.
Here we summarize the main features.  

\vspace{.1in}
\noindent
The full QCD calculation uses 
naive dimensional regularization (NDR) in the $\overline{MS}$ scheme.
A gluon mass $\lambda$
is introduced at intermediate stages of the calculation to regulate 
IR divergences.  This is permissible here since no non-Abelian vertices 
appear in this one-loop calculation. 
A one-loop calculation in full QCD finds, upon expanding through 
\order($1/M$), that the following three current operators are required 
in the 
lattice effective theory:
\begin{eqnarray} \label{jop}
J_L^{(0)} &=& \bar{q} \gamma_5 \gamma_0 Q,  \nonumber \\
J_L^{(1)} &=& - \frac{1}{2M}\,\bar{q} \gamma_5 \gamma_0
(\gammav \cdot \delv) Q,  \nonumber \\
J_L^{(2)} &=& \frac{1}{2M} \, 
\bar{q}\,( \gammav \cdot \overleftarrow{\nabla})
 \gamma_5
 \gamma_0 Q .
\end{eqnarray}
$J^{(0)}_L$ and $J_L^{(1)}$ correspond to discretized versions of
 the first two terms in 
eq.(\ref{jtree}), 
$J_L^{(2)}$ appears only at one loop. 
After carrying out a one-loop mixing matrix calculation in lattice 
NRQCD, one ends up with the final matching relation between $\langle A_0 
\rangle$ in full QCD and matrix elements evaluated in the lattice
 simulations,
\be \label{ajlat}
{\langle  \, A_0 \, \rangle}_{QCD} = 
 \sum_j C_j \langle \, J_{L}^{(j)} \, \rangle   .
\ee
$C_0$, $C_1$ and $C_2$ are the three matching coefficients. 
  Relation (\ref{ajlat}) is independent of the external
states up to lattice artefacts.

\vspace{.1in}
\noindent
An interesting outcome of the one-loop calculation is that the mixing
between $J_L^{(0)}$ and $J_L^{(2)}$ does not vanish in the limit $M \to
\infty$.  This is because of the generation of a term, at order($a \,
\alpha$), which is a lattice artefact. This term can be absorbed  
into  an \order($\alpha \, (a \Lambda_{QCD})$) discretization 
correction to $J_L^{(0)}$:
\be 
\label{jimp}
J_{L}^{(0)} \quad  \rightarrow \quad J_L^{(0)} + C_A \,J_L^{(disc)} ,
\ee
with 
\be
J_L^{(disc)} = a\, \bar{q}\,( \gammav \cdot \overleftarrow{\nabla})
 \gamma_5 \gamma_0 Q .
\ee
The coefficient $C_A$ is also calculated in \cite{pert1}. 
$J_L^{(disc)}$ is the analogue in heavy-light physics 
of the $a \, \partial_\mu P$ discretization 
correction to the light quark axial current (here $P$ is the pseudoscalar 
density) of Ref.~\cite{alpha}.

\vspace{.1in}
\noindent
The matching coefficients $C_j$ are functions of the coupling $\alpha$ and 
of the heavy quark mass and one needs to specify the precise definitions of 
these parameters.
We use $\alpha_V(q^{\ast})$ of reference \cite{lepmac} as our perturbative 
expansion parameter.
  The scale $q^{\ast}$ is not yet known 
 for this matching calculation, however, 
 a value for the $ M \rightarrow \infty$ 
limit has been calculated in reference \cite{herhill} and gives 
$q^{\ast} = 2.18/a$.  
We shall extract $f_B$ for $q^{\ast}= 1/a$ and $q^{\ast}=\pi/a$ and use 
 the difference between the two as an estimate of
the $\alpha^2$ uncertainty.
  It is also important to 
use a common definition for the heavy 
quark mass in the continuum and lattice calculations.  This can be 
ensured in perturbation theory by employing on-shell quark mass 
renormalization and using the pole mass in both theories. 
After writing everything in terms of the pole mass, 
we convert to the bare lattice mass $M_0$, which is a well 
defined quantity even beyond perturbation theory and the mass 
that appears in our lattice simulations. We use 
one-loop perturbation theory for this conversion.

\vspace{.1in}
\noindent
Once \order($1/M^2$) corrections to the current are included at tree level,
eq.(\ref{ajlat}) is modified to,

\be \label{ajlat2}
{\langle  \, A_0 \, \rangle}_{QCD}
= 
 \sum_{j=0,1,2} C_j \langle \, J_L^{(j)} \, \rangle   
 \; + \; \sum_{j=3,4,5} \langle \, J_L^{(j)} \, \rangle ,
\ee
with
$ J_L^{(j)}\;, \; j=3,4,5$, corresponding to the three 
$1/M^2$ current corrections in eq.(\ref{jtree}).  The matching coefficients 
$C_j$, $j = 0,1,2$, do not include one-loop feedback from these latter 
$1/M^2$ current operators.  Such contributions come in at 
\order($\alpha / M^2$).   Eq.(\ref{ajlat2})
 is the NRQCD/Clover heavy-light axial current we use to extract $f_{PS}$.  
It includes terms through \order($(\Lambda_{QCD}/M)^2$), 
\order($\alpha \, \Lambda_{QCD}/M$), \order($\alpha/(aM)$) and 
\order($\alpha \,(a \Lambda_{QCD}) $). 
 We note that powers of both $1/(aM)$ and $\Lambda_{QCD}/M$ are contained in 
our expressions.  
These ratios are both 
considered small in the NRQCD formalism (in practice 
 $1/(aM)$ is sometimes allowed to become as large as $\sim$1).
  In particular, one does not take 
 the limit $aM \rightarrow 0$. The cutoff and therefore 
momenta allowed cannot become larger than $\sim M$ in a non-relativistic theory.

\vspace{.1in}
\noindent
We end this section with two brief comments.  
The first concerns non-perturbative effects that  arise because we have 
power-law contributions to the matching coefficients that go 
as $\alpha/(aM)^n$.  These terms are not separated out explicitly in our
calculations  
but they are evident if coefficients are plotted as a function of $aM$. 
They are there to cancel unphysical ultra-violet pieces in the 
NRQCD current matrix elements, which do not match full QCD. It is possible 
for the $\alpha/(aM)^n$ terms in, say, 
$C_0$ to multiply non-perturbative lattice errors, $\propto (a \, 
\Lambda_{QCD})^n$, in the matrix element $\langle J^{(0)}_L \rangle$ to 
give contributions that look like physical $(\Lambda_{QCD}/M)^n$ 
corrections to $f_B$.  This limits our ability to extract the physical 
$M$ dependence of $f_B$.  The source of these non-perturbative errors 
is a mis-match between the infra-red physics of lattice NRQCD and 
that of full QCD.  After taking the improvement of eq.(\ref{jimp}) into 
account our $\langle J_L^{(0)} \rangle$ 
is designed to match full QCD through 
\order($\alpha \,a$). So these errors appear at 
\order($ \alpha^2 \, 
(a \Lambda_{QCD})$) or \order($( a \, \Lambda_{QCD})^2$). 
 The leading contribution to $f_B$ from these non-perturbative lattice 
artefacts is then $\alpha/(aM) \; \times \; \alpha^2 \, 
(a \Lambda_{QCD}) = \alpha^3 \; \times  \; \Lambda_{QCD}/M$ or 
 $\alpha/(aM) \; \times \; 
(a \Lambda_{QCD})^2 = \alpha \, (a \Lambda_{QCD}) \; \times  \; 
\Lambda_{QCD}/M$ giving an error in the slope of \order($\alpha^3$) or 
\order($\alpha \, (a \Lambda_{QCD})$).  Both these terms are of the same 
order as higher order corrections not yet included in our 
matching calculations; they do not lead to any 
fundamental difficulties.  Better precision can be attained in 
principle by going to more highly improved actions and current operators. 
Conversely we note that the situation is considerably worse for Wilson 
light quarks where discretization errors in the action mean that the 
currents cannot be improved to the level we have achieved here. Then 
non-perturbative lattice artefacts could induce an $\alpha/(aM) \; \times 
\; (a \Lambda_{QCD}) = \alpha \, (\Lambda_{QCD} / M)$ error in $f_B$.  

The second comment concerns cancellation of the power-law contributions 
referred to above.  
For instance, our matrix elements will
contain unphysical contributions that behave as $\alpha/(aM)$ and which are 
cancelled by the perturbative matching procedure. This might be a delicate 
operation if the terms behaving as $\alpha/(aM)$ in the matching 
coefficients and in the matrix elements were much larger than the physical 
terms of \order($\Lambda_{QCD} / M$). In the region in which we work 
none of the perturbative 
coefficients of the $\alpha/(aM)$ terms are unduly large.  No delicate 
cancellation is necessary.  Furthermore the current matrix elements are 
clearly dominated by their physical components, in particular the matrix 
elements of the higher dimension operators, $J_L^{(1)}$ and 
$J_L^{(2)}$, show no sign of  significant $\alpha/(aM)^n$ terms.  This 
will be discussed in more detail elsewhere \cite{scalepaper}.
In our calculations power-law terms have been subtracted perturbatively 
through \order($\alpha/(aM)$).  Uncancelled contributions would come in 
at \order($\alpha^2/(aM)$) and \order($\alpha/(aM)^2$).  There is no 
indication from the behavior of our matrix elements that such 
contributions are particularly large.  Any uncertainty arising from  
them  will be covered by the 
systematic errors that we assign in the next section to higher order
 perturbative and relativistic corrections.

\section{Results }

\vspace{.2in}
\noindent
Details of our data analysis for decay constants are described in 
\cite{othernrqcd,stlouis} and will not be repeated here.
 We start with the discussion 
of 
the size of the various terms in the lattice current defined in 
eq.(\ref{ajlat2}).
Table~\ref{tab:two} summarizes our results for 
\be 
\label{fsqrtmdef}
 f^{(j)}\,\sqrt{M_{PS}} = \frac{1}{\sqrt{M_{PS}}} \, 
 \langle 0|\, J_L^{(j)} \,|PS \rangle, \qquad (j=0\ldots 5),
\ee
evaluated at zero 3-momentum. The light quark has been 
extrapolated/interpolated  linearly to 
$\kappa_l$ or $\kappa_s$.  The matrix elements of $J_L^{(2)}$ are not
shown as they are identical to $J_L^{(1)}$ at zero 3-momentum. The
data show that around the physical $B$ ($aM_0 = 2.22(11)$) the matrix
elements of $J_L^{(1)}$ are $\sim 12\%$ of $J_L^{(0)}$. 
The analogous ratio for the sum of the three $1/M^2$ currents (which partially
cancel) is $\sim 3\% $. It is therefore reasonable to assume that the 
 neglected $1/M^3$ and higher terms are $\leq 1\%$. 

\begin{table}
\begin{center}
\begin{tabular}{|l|l|l|l|l|l|}
\hline
\multicolumn{1}{|c}{$aM_0$} &
\multicolumn{1}{|c}{$a^{3/2}f^{(0)}\sqrt{M}$} &
\multicolumn{1}{|c}{$a^{3/2}f^{(1)}\sqrt{M}$} &
\multicolumn{1}{|c}{$a^{3/2}f^{(3)}\sqrt{M}$} &
\multicolumn{1}{|c}{$a^{3/2}f^{(4)}\sqrt{M}$} &
\multicolumn{1}{|c|}{$a^{3/2}f^{(5)}\sqrt{M}$} \\
\hline
\multicolumn{1}{|c}{} &
\multicolumn{4}{|c|}{$\kappa_{l}$} \\
\hline
1.6 & 0.164(7)    & $-0.0245(12)$ & $-0.0053(4)$ & 0.00363(15) & $-0.0050(3)$  \\
2.0 & 0.167(9)    & $-0.0206(10)$ & $-0.0036(2)$ & 0.00216(10) & $-0.00325(19)$ \\
2.7 & 0.170(11)   & $-0.0162(10)$ & $-0.00220(14)$ & 0.00109(6)  & $-0.00181(13)$ \\
4.0 & 0.174(11)   & $-0.0113(8)$  & $-0.00111(8)$ & 0.00045(3)  & $-0.00082(5)$  \\
7.0 & 0.191(8)    & $-0.0075(4)$  & $-0.00049(2)$ & 0.000144(8) & $-0.000287(15)$ \\
10.0& 0.198(9)    & $-0.0055(2)$  & $-0.000272(12)$ & 0.000072(5) & $-0.000145(8)$  \\
\hline
\multicolumn{1}{|c}{} &
\multicolumn{4}{|c|}{$\kappa_s$} \\
\hline
1.6   & 0.194(5) & $-0.0278(8)$  & $-0.00635(17)$ & 0.00420(11)  & $-0.00576(17)$  \\
2.0   & 0.198(6) & $-0.0236(7)$  & $-0.00446(15)$ & 0.00253(7)   & $-0.00375(12)$   \\
2.7   & 0.205(7) & $-0.0187(6)$  & $-0.00276(10)$ & 0.00129(4)   & $-0.00210(8)$   \\
4.0   & 0.212(8) & $-0.0134(5)$  & $-0.00144(6)$ & 0.00054(2)   & $-0.00098(3)$     \\
7.0   & 0.230(6) & $-0.0088(2)$  & $-0.000620(17)$ & 0.000171(5)  & $-0.000341(10)$     \\
10.0  & 0.237(6) & $-0.0065(17)$  & $-0.000344(9)$ & 0.000083(3)  & $-0.000171(5)$    \\
\hline
\end{tabular}
\end{center}
\caption{Tree-level matrix elements, in lattice units. 
$\;f^{(j)}$ is defined in  eq.\protect(\ref{fsqrtmdef}).
}
\label{tab:two}
\end{table}

\vspace{.2in}
\noindent
The results for $a^{3/2} \, f_{PS}\,\sqrt{M_{PS}}$ with the one-loop matching 
coefficients are summarized in Table 3. 
  We show results for
three values of $\alpha$: 0, $\alpha_V(q^*=1/a)$ and
$\alpha_V(q^*=\pi/a)$. 
  Two of the matching coefficients, $C_0$ and 
$C_1$, include a term $ \alpha \; \ln(aM)/\pi$. 
 In Table 3 we have used the dimensionless 
bare heavy quark mass, $aM_0$, for the argument of these logarithms.  There 
are no problems with large logarithms around the physical $B$ meson, however, 
if one were to work at large $aM_0$, resumming the logarithms would
be more appropriate.  
At the $B$ meson,  the one-loop correction decreases the decay constant by 
$\sim 9  -  14\%$ depending on $q^*$  of which $\sim 3 - 5\%$ are due to  
the one-loop contributions from $J_L^{(1)}$, $J_L^{(2)}$ and 
$J_L^{(disc)}$.  
The $\sim 5$\% uncertainty coming from the variation in $q^*$ will be
taken as a measure of the uncertainty coming from higher order
perturbative corrections.  This is a conservative assignment of 
$\sim 50$\% of the one-loop correction.
The perturbative matching coefficients employed in this article used 
the full $1/M^2$ action of  eq.(\ref{nrqcdact}) with $\delta  H$ given 
by eq.(\ref{deltaH}) (with, however, four-leaf $\Ev$ fields).  
  Matching coefficients from 
an \order($1/M$) NRQCD action (with 
$\delta H 
= - \frac{g}{2\Mbz}\,\sigmav\cdot\Bv $) are also available \cite{pert1} and 
lead to a difference in $f_B$ of about $\sim 1$\%. 
 This comparison gives us some idea 
of the effect from \order($\alpha / M^2$) contributions to matching 
coefficients on $f_B$.  We will assign a $\sim 4$\% error to the full 
\order($\alpha/M^2$) corrections  from  the $1/M^2$ current operators.  This is 
the  size of $\alpha/(aM_0)^2$ and is more conservative than using
the tree level value of the $1/M^2$ corrections.
Adding this in quadrature to the $\sim 5$\% 
\order($\alpha^2$) and $\sim 1$\% \order($1/M^3$) errors gives 
a $\sim 6$\% systematic error from higher order perturbative and relativistic 
corrections.  The dominant discretization errors in the present 
decay constant 
calculation are of \order($(a\Lambda_{QCD})^2$), giving an additional 
systematic error of $\sim 4$\%.

\begin{table}
\begin{center}
\begin{tabular}{|l|l|l|l|l|l|l|}
\hline
\multicolumn{1}{|c}{} &
\multicolumn{3}{|c|}{$\kappa_{l}$} &
\multicolumn{3}{|c|}{$\kappa_s$} \\
\hline
\multicolumn{1}{|c}{$aM_0$} &
\multicolumn{1}{|c}{tree-level} &
\multicolumn{1}{|c}{$q^\ast=\pi/a$} &
\multicolumn{1}{|c|}{$q^\ast=1/a$} &
\multicolumn{1}{|c}{tree-level} &
\multicolumn{1}{|c}{$q^\ast=\pi/a$} &
\multicolumn{1}{|c|}{$q^\ast=1/a$} \\
\hline
1.6 &0.133(7)  & 0.121(6) & 0.114(6) &0.158(5) & 0.144(4) & 0.136(4)  \\
2.0 &0.142(8)  & 0.128(7) & 0.121(7) &0.169(5) & 0.153(5) & 0.144(5)  \\ 
2.7 &0.151(10) & 0.135(9) & 0.127(9) &0.182(7) & 0.164(6) & 0.153(6)  \\ 
4.0 &0.162(10) & 0.144(10)& 0.135(9) &0.196(7) & 0.176(6) & 0.164(6)  \\ 
7.0 &0.183(8)  & 0.162(7) & 0.150(7) &0.220(6) & 0.195(5) & 0.181(5)  \\ 
10.0&0.193(9)  & 0.170(8) & 0.158(8) &0.231(6) & 0.204(6) & 0.189(5)  \\ 
\hline
\end{tabular}
\end{center}
\caption{ Decay matrix elements $f\protect\sqrt{M}$ in lattice units. The first
 column 
gives tree-level results, 
the second includes renormalization constants using $\alpha_V$ at $q^\ast = \pi
/a$, and
the third using $\alpha_V$ at $q^\ast = 1/a$.}
\label{tab:renorm}
\end{table}



\vspace{.1in}
\noindent
 Table \ref{tab:fb} shows results for $f_{PS}$ at the physical $b$-quark mass 
 for
several choices of the scale. 
For each scale, $aM_b^0$ denotes the value of the dimensionless 
bare heavy quark mass which gives the correct $B$ meson mass.  
Using the scale $a^{-1}_{m_\rho} = 1.92(7)$ GeV and averaging over the two
 $q^*$'s, 
our final estimate in the quenched approximation is then, 
\begin{eqnarray} \label{fbresult}
f_B &=& 147(11)(^{+8}_{-12})(9)(6) \rm{MeV}, \nl
f_{B_s} &=& 175(08)(^{+7}_{-10})(11)(7)(^{+7}_{-0}) \rm{MeV}.
\end{eqnarray}
The first error comes from statistics plus  
extrapolation/interpolations in $\kappa$ and 
$aM_0$. Scale uncertainties are reflected in the second error with the 
upper and lower limits coming from $a^{-1} = 2.0$ GeV or $a^{-1}=1.8$ GeV 
respectively. The third error is due to  higher order relativistic and 
perturbative corrections and the fourth due to discretization 
corrections.
The central value of $f_{B_s}$ has been calculated 
fixing the strange quark mass from the $K$ meson; the fifth error bar 
comes from the difference in $\kappa_s$ from the $\phi$ and the $K$.

\begin{table}
\begin{center}
\begin{tabular}{|l|l|l|l|l|l|}
\hline
 & &
\multicolumn{2}{|c}{$f_B\;$ [MeV]} &
\multicolumn{2}{|c|}{$f_{B_s}\;$ [MeV]} \\
\hline
$a^{-1}\;$ [GeV] &
$aM_b^0$ &
$q^\ast=\pi/a$ &
$q^\ast=1/a$ &
$q^\ast=\pi/a$ &
$q^\ast=1/a$ \\
\hline
 1.8   & 2.39(7)  & 140(9) & 131(8)  & 171(5) & 160(5)  \\ 
 1.92  & 2.22(6)  & 151(9) & 142(9)  & 180(6) & 169(6)  \\ 
 2.0   & 2.11(6) & 159(9) & 150(9)  & 187(6) & 176(6)  \\
\hline
\end{tabular}
\end{center}
\caption{ Decay constants $f_B$ and $f_{B_s}$
 in MeV. Statistical errors only are shown.
$aM_b^0$ is the dimensionless bare heavy quark mass which leads to  
the correct $B$ meson mass.  
 }
\label{tab:fb}
\end{table}

\vspace{.1in}

\begin{table}
\begin{center}
\begin{tabular}{|l|l|l|l|}
\hline
\multicolumn{1}{|c}{$aM_0$} &
\multicolumn{1}{|c}{tree-level} &
\multicolumn{1}{|c}{$q^\ast = \pi/a$} &
\multicolumn{1}{|c|}{$q^\ast =  1/a$} \\
\hline
1.6 & 1.18(4)(5) & 1.18(4)(5) & 1.19(4)(4) \\
2.0 & 1.19(4)(4) & 1.19(4)(4) & 1.19(4)(5) \\
2.7 & 1.20(5)(5) & 1.20(5)(5) & 1.21(5)(5) \\
4.0 & 1.21(5)(5) & 1.21(5)(5) & 1.21(5)(5) \\
7.0 & 1.19(4)(5) & 1.20(4)(4) & 1.20(4)(4) \\
10.0& 1.19(4)(4) & 1.19(4)(5) & 1.19(4)(5) \\
\hline
\end{tabular}
\end{center}
\caption{$f_{B_s}/f_{B}$, without renormalization constants (left),
 $q^\ast = \pi/a$ (middle), and $q^\ast = 1/a$. 
}
\label{tab:fBs/fBd}
\end{table}

\vspace{.1in}
\noindent
Another quantity of interest is the ratio  $f_{B_s}/f_{B}$. 
It can be determined to a greater accuracy than $f_B$ and $f_{B_s}$
 separately,
since the direct
dependence on the lattice scale drops out, and the statistical errors
partly cancel. We obtain this ratio for 
each value of the heavy mass using,
\be
\frac{f_{B_s}}{f_{B}} = \frac{(f\sqrt{M})_{\kappa = \kappa_s}}
{(f\sqrt{M})_{\kappa = \kappa_l}} \; \frac{\sqrt{M_B}}{\sqrt{M_{B_s}}} ,
\ee
where $M_{B}$ and  $M_{B_s}$ are the experimental meson masses.
Table \ref{tab:fBs/fBd} shows our results.
One sees that
 one-loop renormalization has negligible effect on $f_{B_s}/f_{B}$. This also
indicates that the errors from higher orders in perturbation theory are 
negligible. 
 In the range studied, the data do not show any significant dependence 
 on $a M_0$. 
 Linearly interpolating to $aM_b^0$, our best
estimate of the ratio is
\be \label{fbratio}
\frac{f_{B_s}}{f_{B}} = 1.20(4)(^{+4}_{-0}) .
\ee
The first error is statistical and 
the second error comes from the uncertainties in fixing $\kappa_s$.

\vspace{.1in}
\noindent
Eqs.~(\ref{fbresult}) and (\ref{fbratio}) give the main results of 
this article and represent the most complete results to date for $f_B$ and 
$f_{B_s}/f_B$ using NRQCD $b$-quarks. It is interesting to compare them 
with other calculations that have used a relativistic formalism for 
the heavy $b$-quark.  
 Table~\ref{tab:comparison} gives such a comparison with the APE \cite{APE}, 
Fermilab \cite{fermilab}, JLQCD \cite{JLQCD}, MILC \cite{MILC} and 
UKQCD \cite{UKQCD} collaborations.  Details of the simulation parameters 
and analyses are different among these groups (for recent reviews see 
 \cite{review}).  For example MILC uses the Wilson quark action, UKQCD and 
APE results are for the clover action with $c_{SW} = 1$, Fermilab  
uses tree-level tadpole improved clover and JLQCD uses one-loop tadpole 
modified clover. Only JLQCD, MILC and Fermilab  extrapolate
to $ a = 0$, and JLQCD and Fermilab include $M$-dependent renormalization 
constants at \order($\alpha$). They do not take into account 
one-loop mixing with $1/M$ current operators or the discretization 
correction that arises at \order($\alpha(a \Lambda_{QCD})$). We expect that at 
comparable  $\beta$ and $M$ these two corrections  should be of similar size 
in their formalism as we find here ($\sim 3-5\%$).  
 In spite of the range of approaches and  approximations used,
it is, nevertheless, encouraging to see that very 
different formulations of heavy quarks give results for $f_B$ and $f_{B_s}$
 that are consistent within errors.  The ratio $f_{B_s}/f_B$, in which 
many of the errors cancel, shows a larger spread among the groups. The NRQCD 
number lies $1 \sim 2 \sigma$ higher than the more recent relativistic 
heavy quark results in Table~\ref{tab:comparison} by APE, Fermilab and MILC.

\begin{table}
\begin{center}
\begin{tabular}{|c|c|c|c|}
\hline
\multicolumn{1}{|c}{Collaboration} &
\multicolumn{1}{|l|}{$f_B$[MeV]}   &
\multicolumn{1}{|l|}{$f_{B_s}$[MeV]}   &
\multicolumn{1}{|l|}{$f_{B_s}/f_B$} \\
\hline
This work                              & $147(^{+17}_{-20})$ & $175(^{+
18}_{-18})$  & $1.20(^{+6}_{-4})$ \\
\hline
APE~\protect\cite{APE}                 &  180(32)            & 205(35)  & 
1.14(8)   \\
Fermilab~\protect\cite{fermilab}       & $ 164(^{+16}_{-14})$           
 & $185(^{+16}_{-11})$   & $1.13(^{+5}_{-4})$  \\
JLQCD~\protect\cite{JLQCD}             &  163(16)               &  175(18)
 &     \\
MILC~\protect\cite{MILC}               & $153(^{+40}_{-16})$    & $164(^{+5
0}_{-16})$ & 1.10(6)   \\
UKQCD~\protect\cite{UKQCD}              & $160(^{+53}_{-20})$ & 
$194(^{+62}_{-10})$ & 1.22(4) 
  \\
\hline
\end{tabular}
\end{center}
\caption{Comparison with quenched results for $f_B$, $f_{B_s}$ and 
$f_{B_s}/f_B$ from recent lattice 
calculations using relativistic quarks. Errors include statistical and
systematic errors, added in quadrature.}
\vspace{.1in}
\noindent
\label{tab:comparison}
\end{table}

\vspace{.1in}
\noindent
Our $B$ meson decay constant results 
 are still within the quenched approximation and at a fixed 
value of the lattice spacing.  Calculations are underway to address these 
sources of uncertainty.  The first results from $n_f=2$ dynamical 
configurations (without the $1/M^2$ tree-level corrections) show 
that, within current statistical and systematic errors, 
 unquenching effects are not yet visible~\cite{sgo}. 
Higher statistics and smaller dynamical quark masses are required 
to resolve this issue. 
Since we are using improved actions and currents we expect scaling 
violations to be small in quenched calculations at $\beta = 6.0$. 
  Further quenched simulations at 
other lattice spacings ($\beta = 5.7$ and $\beta = 6.2$) have been 
initiated to check this. 
Preliminary results at $\beta= 5.7$ on UKQCD configurations
are encouraging~\cite{hein}, however, 
to make any definitive statement requires a full analysis at all three 
$\beta$ values.

\section{Summary}

This article describes a quenched calculation of $B$ meson decay constants, 
$f_B$ and $f_{B_s}$, using NRQCD $b$-quarks and clover light quarks. 
 Tree-level 
corrections to the heavy quark action and the heavy-light currents have been 
included through \order($1/M^2$).  The full \order($\alpha/M$) matching 
between continuum QCD and lattice NRQCD currents has been incorporated 
together with an \order($a \, \alpha$) discretization correction to 
the heavy-light axial current.  This is the first time both mass-dependent 
matching factors and the \order($a \, \alpha$) current correction have been 
included in a lattice determination of $f_B$.  Our final results for 
$f_B$, $f_{B_s}$ and $f_{B_s}/f_B$ are given in 
eqs.~(\ref{fbresult}) and 
(\ref{fbratio}).  
 The next step in our program 
is to study unquenching, scaling and finite volume effects, in addition 
to improving statistics.  Work is already in progress on these fronts. 

\vspace{.2in}
\noindent
\underline{Acknowledgements}

\vspace{.1in}
\noindent
This work has been supported by grants from the US Department of Energy, 
(DE-FG02-91ER40690, DE-FG03-90ER40546, DE-FG05-84ER40154, 
DE-FC02-91ER75661, DE-LANL-ERWE161),
NATO (CRG 941259), PPARC and the Royal Society of Edinburgh.  A. Ali Khan 
is grateful to the Graduate School of the Ohio State University for 
a University Postdoctoral Fellowship. 
C. Davies thanks the Institute for Theoretical Physics, Santa Barbara, for
hospitality and the Leverhulme Trust and Fulbright
Commission for funding while this work was being completed.
J. Sloan would like to thank the Center for Computational 
Sciences, University of Kentucky, for support. 
We thank Joachim Hein, Peter Lepage and Tetsuya Onogi for useful 
discussions.  
  The simulations reported here
were done on CM5's at the Advanced Computing Laboratory at Los Alamos
under a DOE Grand Challenges award, and at NCSA under a Metacenter
allocation.

\end{document}